\documentclass{aa}
\usepackage{epsfig}
\usepackage{graphicx}
\begin{document}

\title{Two emission line objects with $z>0.2$ in the optical
filament apparently connecting the Seyfert galaxy
NGC 7603 to its companion} 

\author{
M. L\'opez-Corredoira\inst{1,2} \and Carlos M. Guti\'errez\inst{1}}
\institute{$^1$ Instituto de Astrof\'\i sica de Canarias, E-38200 La Laguna, 
Tenerife, Spain\\
$^2$ Astronomisches Institut der Universit\"at Basel,
Venusstrasse 7, CH-4102 Binningen, Switzerland
}

\offprints{martinlc@astro.unibas.ch}

\date{Received xxxx / Accepted xxxx}

\abstract{
We present new spectroscopic observations of an old case of anomalous 
redshift---NGC 7603 and its companion.
The redshifts of the two galaxies which are apparently
connected by a luminous filament are $z=0.029$ and $z=0.057$
respectively. We show that in the luminous filament there
are two compact emission line objects with $z=0.243$ and $z=0.391$.
They lie exactly on the line traced by the filament connecting the galaxies.
As far as we are aware, this is the most impressive 
case of a system of anomalous redshifts discovered so far.
}
\titlerunning{NGC 7603}

\maketitle

\section{Introduction}

Thirty years ago it was shown that NGC 7603 (Mrk 530, Arp 92)
is a remarkable example of an anomalous redshift association (Arp 1971). 
NGC 7603 is a
Seyfert 1 galaxy with strong spectral variability (Kollatschny et al. 2000). 
A smaller galaxy (NGC 7603B, denoted by object 1 in
the figure 1) lies at the end of a filament which apparently connects both
galaxies. The galaxies have redshifts corresponding to
8700 km/s and 17000 km/s respectively. 

Sharp (1986) suggested that the absence of emission lines in the
smaller galaxy argued against a possible interaction
between NGC 7603 and the smaller companion
(NGC 7603B$\equiv $object 1). However, the non-detection of emission lines
is not proof against the existence of a physical connection.
In interactions and ejections with a
larger galaxy, the gas is often stripped out of a stellar
system; so the lack of emission lines could be taken as an 
indication of interaction rather than non interaction.
It might also be that NGC7603B, although roughly at the same distance, 
is not close enough to NGC7603 to have tidal effects and star formation.

NGC 7603 and other examples,
in which two galaxies with different redshifts
are apparently connected by a filament, have been considered fortuitous by
most researchers. However, we decided that a deep analysis of
this system should be carried out to try to give an answer to
the controversy. We considered it
important to determine the redshifts of the two
observed knots (Arp 1971) (objects 2 and 3) in the filament.
For some time after the discovery of the system, attempts
were made to obtain spectra of the objects in the bridge, but because of
limitations of the equipment available in the 1970s, none was successful
(M. Burbidge, private comm.). 
In this paper we report observations in which we were succesful in obtaining
spectra of not two but four objects connected by the filament.
In \S \ref{.redsh}, we discuss the spectra of objects 2 and 3.

\section{Observations}

Table \ref{Tab:NOT} summarizes the observations\footnote{Based 
on observations made with the Nordic Optical Telescope,
operated on the island of La Palma jointly by Denmark, Finland,
Iceland, Norway, and Sweden, in the Spanish Observatorio del 
Roque de los Muchachos of the Instituto de Astrof\'\i sica de 
Canarias.} used in this paper. 
The image in R-band (Fig. 1a) shows NGC 7603 with 3 companions:
NGC 7603B (object 1), object 2 and object 3. We took long-slit spectra
of the southern part of NGC 7603B, and objects 2 and 3;
all the spectra were obtained 
within the same 5 arcseconds-width slit indicated 
by a dashed line in Fig. 1b.

\begin{table*}[htb] 
\begin{center} 
\caption{Source of NGC7603 data used in this paper: R-band image 
as shown in Fig. 1, and a long-slit spectrum along the dashed lines
marked in Fig. 1b whose extracted spectra are shown in Fig. 2.} 
\begin{tabular}{|c|c|c|} \hline
--- & R-band image & slit-spectrum 4000-7000\AA \\ \hline
Telescope & NOT-2.6 m. (La Palma-Spain) & 
 NOT-2.6 m. (La Palma-Spain) \\
Instrument & ALFOSC & ALFOSC/grism 4, aperture: $5''$ \\
Resolution & $0.188''$/pixel & $0.188''$/pixel; 2.96 \AA/pixel \\
Date & 2000, June 13th & 2001, August 12th \\
Exposure time/Moon & 900 s./dark & 14225 s./grey-dark \\ \hline 
\end{tabular} 
\end{center} 
\label{Tab:NOT}
\end{table*} 

\begin{table*}[htb] 
\begin{center} 
\caption{Characteristics of the observed objects.} 
\begin{tabular}{|c|c|c|c|c|c|} \hline
Object & Spectral lines & structure & Eq. coordinates (J2000) & magnitude & redshift (heliocentric) \\ \hline
1 (NGC 7603B) & absorption & extended & $\alpha =23^{\rm h}19^{\rm m}00.1^{\rm
s}$, $\delta
=+0^\circ14'7''$ & $m_B=16.8$ $^{1}$ & $0.058\pm 0.002$ \\ 
2 & emission & point-like & $\alpha =23^{\rm h}18^{\rm m}59.4^{\rm s}$, $\delta
=+0^\circ14'4''$ & $m_R=21.8\pm 0.2$ & $0.243\pm 0.001$ \\
3 & emission & point-like & $\alpha =23^{\rm h}18^{\rm m}57.7^{\rm s}$, $\delta
=+0^\circ14'2''$ &
$m_R=21.4\pm 0.2$ & $0.391\pm 0.001$ \\
Filament & absorption & extended & from NGC 7603 to NGC 7603B  & $\sim $ 
23.5/arcsec$^2$ & 
$0.030 \pm 0.001$ \\ \hline 
\end{tabular} 
\end{center} 
\label{Tab:objetos}
{\footnotesize $^1$ Sharp (1986)}
\end{table*} 

\begin{figure}[!]
\begin{center}
\vspace{1cm}
\mbox{\epsfig{file=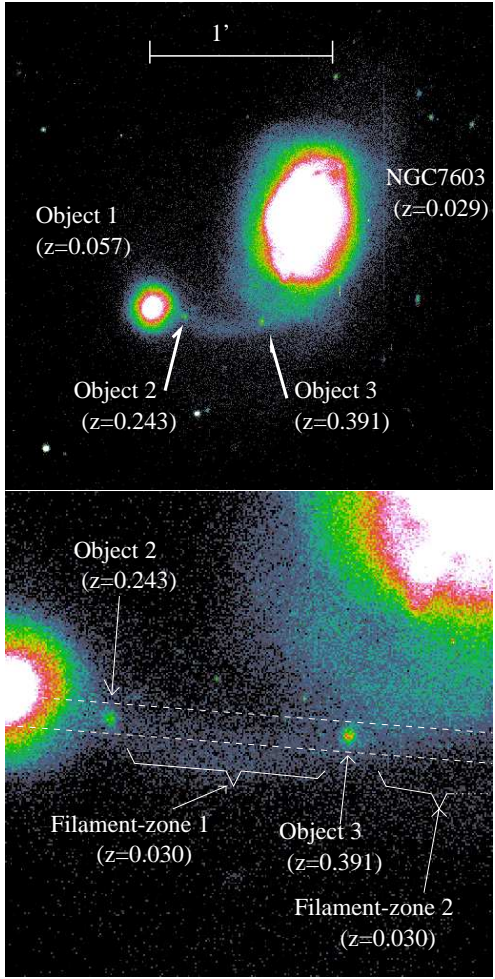,height=13cm}}
\end{center}
\caption{a) NGC7603 in R-band taken with the NOT-2.6 m. 
telescope (La Palma). Four objects with different 
redshift, plus a filament apparently connecting all of them, were observed.
b) Magnification of Fig. 1a); dashed lines indicate the position of the 
long-slit where the spectra of the objects were taken. 
This figure summarizes the content of the paper: the case of a galaxy
(NGC 7603) with 3 companions, all of them with different redshift.
The filament between NGC 7603 and NGC 7603B (object 1) shows clearly. 
A knot (object 2) is centered in the line of the filament
and positioned where the filament connects to NGC7603B. The other
knot (object 3) is also centered in the line of the filament,
and is positioned where the filament connects with NGC7603.
The astonishing fact comes not from this image itself but
from Fig. 2, which gives the redshifts of the objects.
Everything points the four objects being connected among themselves,
but how to explain the different redshifts? Or, in case all of them
have different distances, how to explain that their projections in the sky
give this extremely low probable configuration?}
\label{fig1}
\end{figure}

\section{Discrepant redshifts}
\label{.redsh}

Figures 1a,1b show clearly
the filament between NGC 7603 and NGC 7603B (object 1). A knot
(object 2) is perfectly centered in the line of the filament
and positioned where the filament connects to NGC7603B. The other
knot (object 3) is also perfectly centered to within 1 arcsecond in 
the filament, and is positioned where the filament connects with NGC7603.
There is also a second filament which sweeps around from the main galaxy through the
position of the companion NGC7603B (Arp 1971, 1975). The halo of the
common system extends over the northern area of NGC7603B, that is
at the left side of the center of NGC7603 in Fig. 1a. However,
the halo does not extend on the right position in the same way.

\begin{figure}[!h]
{\par\centering \resizebox*{6.7cm}{6.7cm}{\epsfig{file=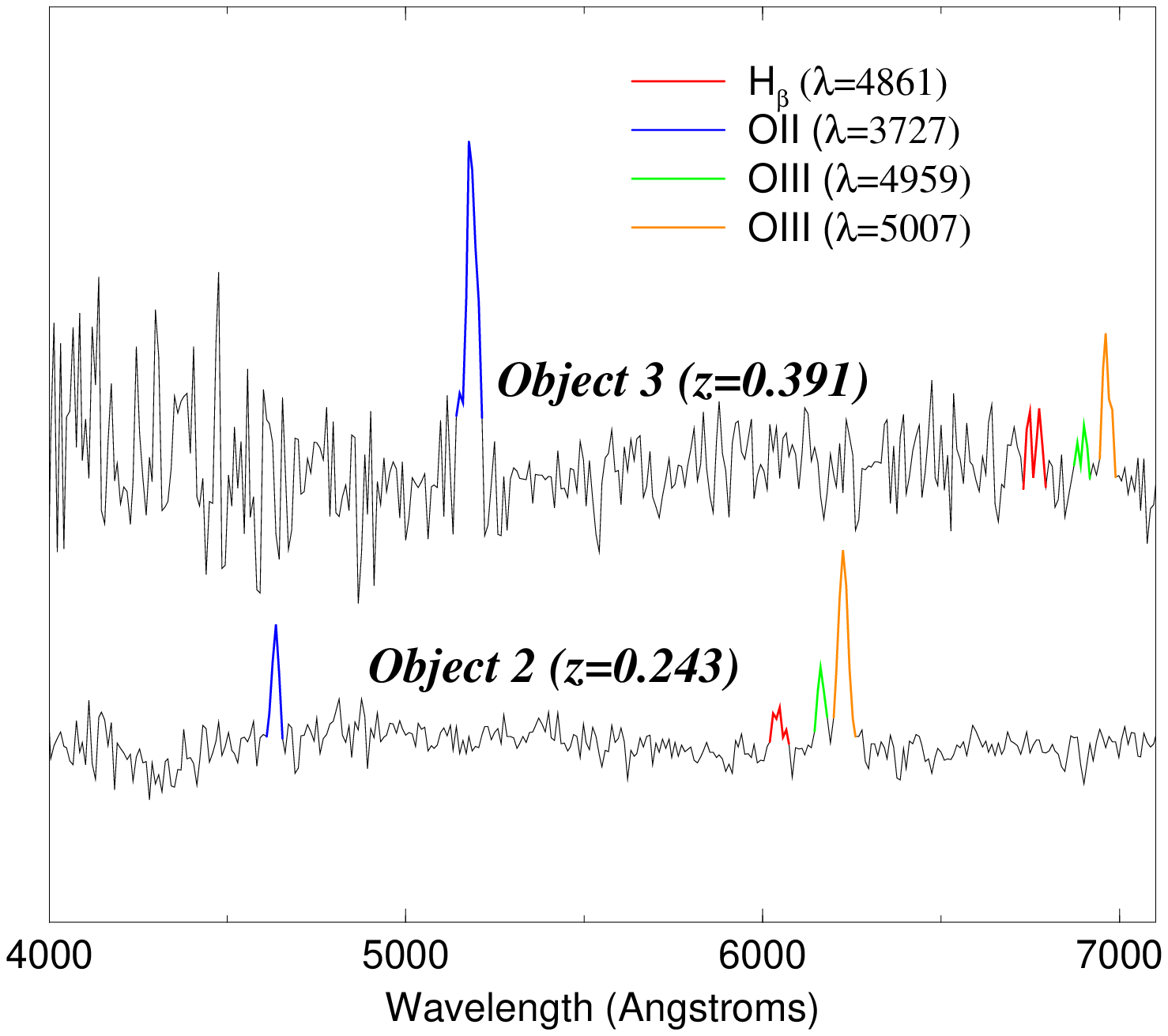,height=6.7cm}}
\resizebox*{6.7cm}{6.7cm}{\epsfig{file=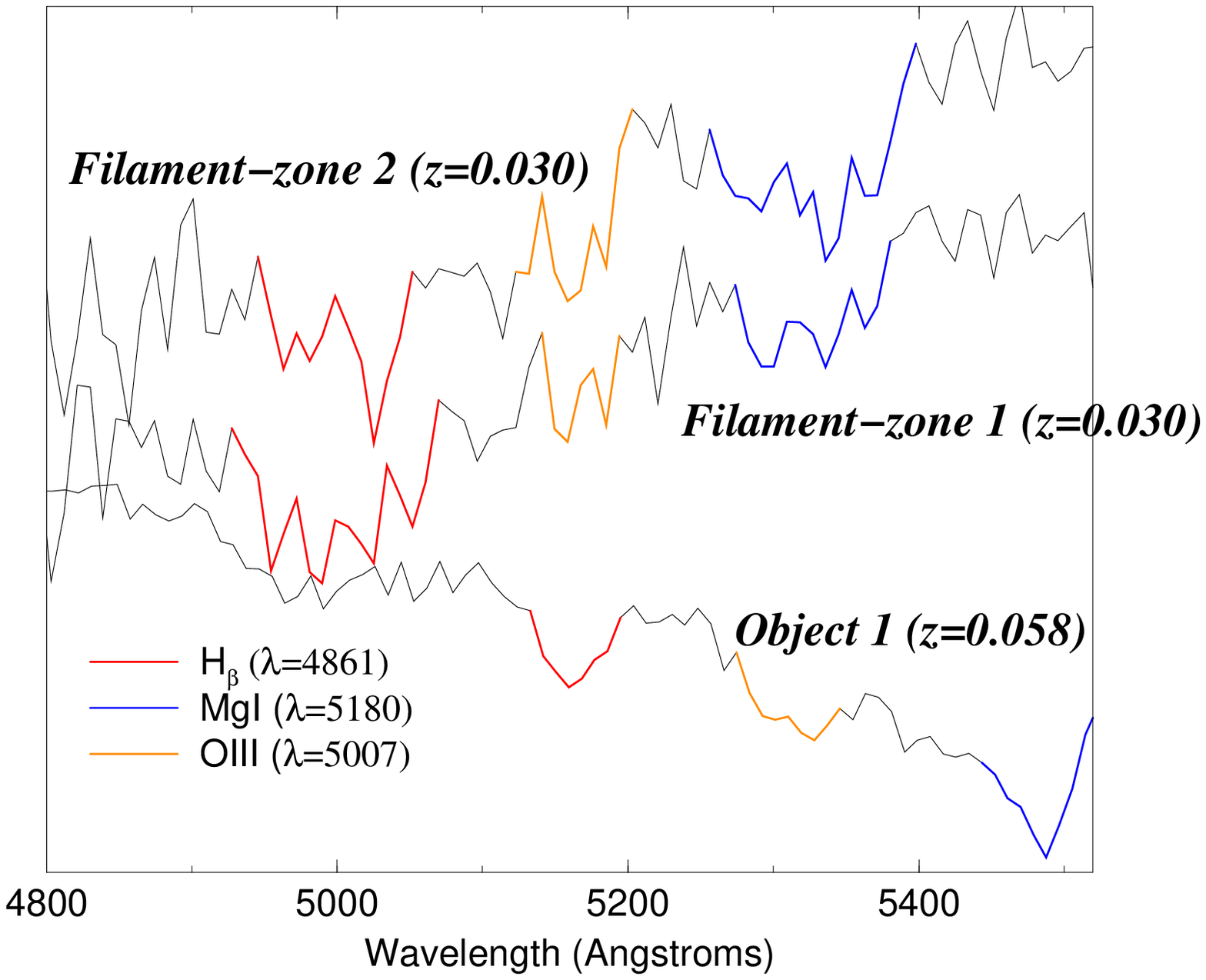,height=6.7cm}}\par}
\caption{a) Spectra in the range 4000-7000 \AA \ of the objects 2 
and 3 (see Fig. 1). OII, OIII and H$_{\beta}$ are emission lines
identified in them. 
b) Spectra in the range 4800-5500 \AA \ of
NGC 7603B and the filament in two different zones (see Fig. 1).
Here we show the lines H$_\beta $, OIII (5007 \AA ), MgI in absorption.
NGC 7603B has again a discordant redshift with respect NGC 7603
(already known from previous work) while the filament, with a much
lower signal/noise, in both regions is presumably
consistent with a redshift similar to that of NGC 7603.
All the spectra were taken with the NOT-2.6 m. 
telescope (La Palma), continuum normalized and 
binned by a factor 3 for better signal/noise ratio, so the resolution
in these plots is 8.9 \AA/pixel.}
\label{fig2}
\end{figure}

We have determined the redshift of the objects 1(NGC 7603B) to 3.
Table \ref{Tab:objetos} summarizes the information
about these objects.
Fig. 2a shows the spectra of the objects 2 and 3, which are emission line
sources with redshifts 0.243 and 0.391 respectively. OII, H$_\beta$ and
the OIII doublet are plotted in Fig. 2a. They can be classified as 
broad line objects (Seyfert 1/quasar)
since the H$_\beta$ line in both cases has a $FWHM\approx 2400$ km/s,
broader than the forbidden lines OIII with a $FWHM\approx 1500$ km/s 
(the width of the narrower lines is mainly instrumental),
which implies an intrinsic broadening of H$_\beta \approx 1900$ km/s.
Seyfert galaxies and quasars are basically the same, and differ only in the 
proportion of light coming from the active nucleus and the host galaxy,
so we do not make a distinction between these objects; the important
feature is the broadening of $H_\beta $.
Fig. 3 gives a zoom of Fig. 2a for the object 2 
around $H_\beta $ and OIII lines which shows better 
the broadening of H$_\beta$. For the object 3, the broadening
is similar but the signal/noise ratio is smaller and perhaps not
conclusive. A high contrast between the widths of the narrow
and the wide lines cannot be expected because the aperture
used in our slit produce a large instrumental broadening of the lines.
These spectra are similar to other spectra for quasars/Seyferts 1
in other examples of anomalous redshifts (Burbidge 1995, 1997).
Moreover, they are point-like objects (FWHM equal
to the seeing: $\approx 1.0''$) which is another expected feature in
this kind of objects.
Other authors (Burbidge 1995, 1997; Arp et al. 2001) have also 
reported the detection of quasars/Seyferts 1 apparently ejected by 
a parent Seyfert galaxy.

If we did not trust either the argument of the broadening of H$_\beta $ 
nor the argument of their being point-like objects, they would be narrow 
emission line extended objects. They would be HII galaxies or LINERs
because $R_{23}\equiv \frac{[OII]+[OIIIa]+[OIIIb]}{H_\beta}=$4.2 and
5.5 respectively for objects 2 and 3, while
Seyfert 2 should have this ratio larger than 12.5
(Dessauges-Zavadsky et al. 2000). They should not be LINERs
because $\frac{[OIIIb]}{H_\beta}=2.4$ and 1.0 respectively for objects 2 and 3, 
larger than 0.5 (Filippenko \& Terlevich 1992).
Hence, they would be HII galaxies. 

\begin{figure}
\begin{center}
\vspace{1cm}
\mbox{\epsfig{file=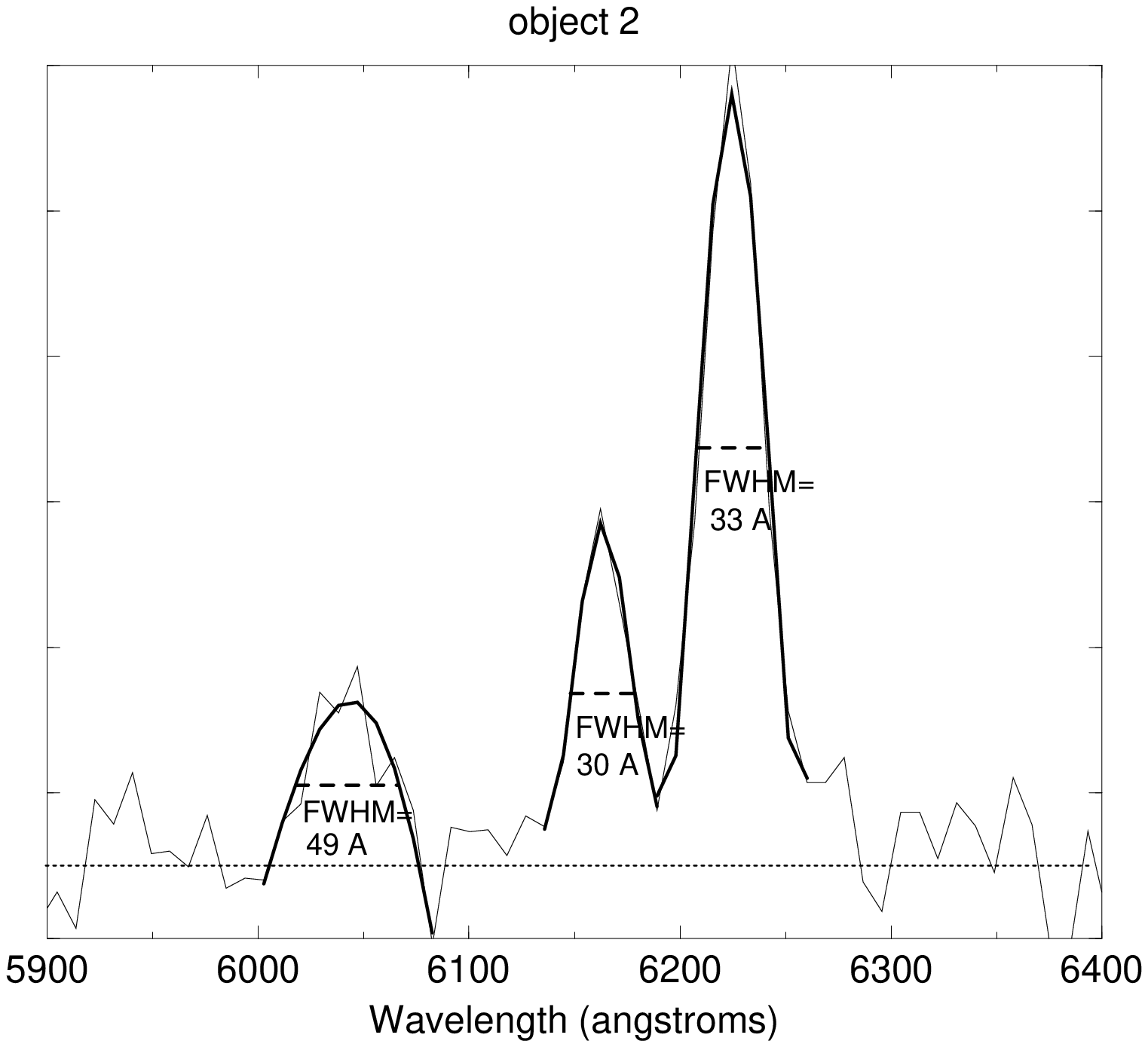,height=6.7cm}}
\end{center}
\caption{Zoom of Figure 2 a) for object 2. Note the excess broadening
of H$_\beta $.}
\label{fig3}
\end{figure}

NGC 7603B and the filament
present absorption lines, respectively with redshifts 0.058 and 0.030 
(like NGC7603). In Fig. 2b, it is shown a part of the spectra, where
we can identify H$_\beta$, OIII and MgI. The filament zones were taken as
a sum of 87 columns (0.188 arcseconds each) 
in the zone 1 and 173 columns in the zone 2.
Although the identification of
lines for the filament is not as clear as for the emission line
objects (because the signal/noise ratio is lower), we can tentatively
attribute to it a redshift of 0.030 consistent with the NGC7603 redshift. 
This means that we do not see a progressive change of the redshift
between 0.029 and 0.057, which would be expected if both 
galaxies were at the same distance and the different redshift were
due to a Doppler effect of peculiar motions.
This is, in some way, the least problematic of our measures. It would be
ideal to have longer exposures or a larger telescope to confirm this
measurement, although in any case, the most remarkable fact here is
the redshift of objects 1(NGC 7603B)-3 rather 
than the redshift of the filament.

NGC 7603B is a galaxy with magnitude $m_B=16.8$ (Sharp 1986) and we 
have measured magnitudes $m_R=21.8$ and $m_R=21.4$ respectively for 
objects 2 and 3 (table \ref{Tab:objetos}). Those two objects are
more prominent in blue plates than in red plates (Sharp 1986);
object 2 is even visible in the blue POSSII plates (with limiting magnitude
between 21.0 and 21.5), so we can be sure that the equivalent $m_{b_j}$ 
is not fainter than $\approx 21.9$, even brighter if there were
some extinction through the filament. 
Up to these magnitudes, we have $N_1\sim 8$ deg$^{-2}$
(from complete galaxy counts: Metcalfe et al. 1991); and
$N_2\sim N_3\sim 70-350$ deg$^{-2}$, depending on the classification
of the objects: $\sim 70$ if they were quasars/Seyfert 1 
(from complete quasar counts: Boyle et al. 1991), 
or 5 times more if they were HII galaxies (10\% of
the emission line objects are quasars/Seyfert 1 and
50\% are HII galaxies; Ho et al. 1997, Meyer et al. 2001).
That is, there should be one object like these per each square
of 3-7 arcminute size (20 arcminute size for NGC 7603B); much
larger than the area of the filament ($\sim 100$ arcsec$^2$).

\section{Discussion and conclusions}
\label{.prob}

We have clearly shown that two of the compact emission lines
objects in the filament have redshifts very much greater than those of
NGC 7603 and its companion galaxy. 
Thus we have presented a very well known
system with anomalous redshifts, NGC 7603,
to be an apparently much more anomalous than was previously thought.
There are 4 objects with very different redshifts apparently
connected by a filament associated with the lower redshift galaxy. 
This system is at present the most spectacular case that we know
among the candidates for anomalous redshift.
Future studies of this system are clearly warranted.

Acknowledgments: 
We gratefully acknowledge the anonymous referee for helpful comments.
Thanks are also given to Victor P. Debattista and Gustav Tammann 
(Astron. Inst. Basel) for helpful discussion about the present paper.


\begin{thebibliography}{99}

\bibitem{} Arp H., 1971, Astrophys. Lett. 7, 221

\bibitem{} Arp H., 1975, PASP 87, 545

\bibitem{} Arp H., Burbidge E. M., Chu Y., Zhu X., 2001, 
ApJ 553, L11 

\bibitem{} Boyle B. J., Jones L. R., Shanks T. A., 1991, MNRAS 
251, 482

\bibitem{} Burbidge E. M., 1995, A\&A 298, L1

\bibitem{} Burbidge E. M., 1997, ApJ 484, L99


\bibitem{} Dessauges-Zavadsky M., Pindao M., Maeder A., Kunth D., 2000,
A\&A 355, 89

\bibitem{} Filippenko V. A., Terlevich R. J., 1992, ApJ 397, L79

\bibitem{} Ho L. C., Filippenko A. V., Sargent W. L. W., 1997, ApJ 487, 568

\bibitem{} Kollatschny W., Bischoff K., Dietrich M., 2000,
ApJ 361, 901

\bibitem{} Metcalfe N., Shanks T., Fong R., Jones L. R., 1991,
MNRAS 249, 498

\bibitem{} Meyer M. J., Drinkwater M. J., Phillips S., Couch W. J.,
2001, MNRAS 324, 343

\bibitem{} Sharp N. A., 1986, ApJ 302, 245

\end{thebibliography}
\end{document}